\documentclass[prx,superscriptaddress,aps,amsmath,amssymb,amsfonts,floatfix,reprint,raggedbottom,longbibliography]{revtex4-2}
        \usepackage{graphicx}
        \usepackage{xr-hyper}
        \usepackage[colorlinks=true,citecolor=blue,linkcolor=blue,urlcolor=blue]{hyperref}
        \usepackage{balance,wrapfig}
        \usepackage{dcolumn}
        \usepackage{xcolor}
        \usepackage{multirow}
        \usepackage{array,booktabs}
        \usepackage[ruled,vlined,linesnumbered]{algorithm2e}
        \usepackage{titlesec}
        \usepackage{lmodern}
        \usepackage{times}
        \usepackage[utf8]{inputenc}
        \usepackage[para]{footmisc}
        \usepackage{mathtools}
        \allowdisplaybreaks
        
        \makeatletter
        \renewcommand{\fnum@figure}{\textbf{Fig.~\thefigure}}
        \makeatother

        
        \newcolumntype{L}{>{\arraybackslash}m{3.9 cm}}
        \newcolumntype{C}{>{\centering\arraybackslash}m{3.2 cm}}
        \newcolumntype{G}{>{\centering\arraybackslash}m{1.41 cm}}
        
\begin{document}
        
        \title{Probabilistic approximate optimization using single-photon avalanche diode arrays}
        
        \author{Ziyad Alswaidan}
        \thanks{These authors contributed equally.}
        \affiliation{Department of Electrical and Computer Engineering, Carnegie Mellon University, Pittsburgh, PA 15213, USA}
        
        \author{Abdelrahman S. Abdelrahman}
        \thanks{These authors contributed equally.}
        \affiliation{Department of Electrical and Computer Engineering, University of California, Santa Barbara, Santa Barbara, CA 93106, USA}

        \author{Md Sakibur Sajal}
        \thanks{These authors contributed equally.}
        \affiliation{Department of Electrical and Computer Engineering, Carnegie Mellon University, Pittsburgh, PA 15213, USA}

        \author{Shuvro Chowdhury}
        \affiliation{Department of Electrical and Computer Engineering, University of California, Santa Barbara, Santa Barbara, CA 93106, USA}
        
        \author{Kai-Chun Lin}
        \affiliation{Department of Electrical and Computer Engineering, Carnegie Mellon University, Pittsburgh, PA 15213, USA}
        
        \author{Hunter Guthrie}
        \affiliation{Department of Electrical and Computer Engineering, Carnegie Mellon University, Pittsburgh, PA 15213, USA}

        \author{Sanjay Seshan}
        \affiliation{Department of Electrical and Computer Engineering, Carnegie Mellon University, Pittsburgh, PA 15213, USA}
        
        \author{Shawn Blanton}
        \affiliation{Department of Electrical and Computer Engineering, Carnegie Mellon University, Pittsburgh, PA 15213, USA}
        
        \author{Flaviano Morone}
        \affiliation{Center for Quantum Phenomena, Department of Physics, New York University, New York, NY 10003 USA}
        
        \author{Marc Dandin}
        \altaffiliation{Email: \texttt{mdandin@andrew.cmu.edu}, \texttt{camsari@ucsb.edu}, \texttt{tsrimani@andrew.cmu.edu}}
        \affiliation{Department of Electrical and Computer Engineering, Carnegie Mellon University, Pittsburgh, PA 15213, USA}
        
        \author{Kerem Y. Camsari}
        \altaffiliation{Email: \texttt{mdandin@andrew.cmu.edu}, \texttt{camsari@ucsb.edu}, \texttt{tsrimani@andrew.cmu.edu}}
        \affiliation{Department of Electrical and Computer Engineering, University of California, Santa Barbara, Santa Barbara, CA 93106, USA}
        
        \author{Tathagata Srimani}
        \altaffiliation{Email: \texttt{mdandin@andrew.cmu.edu}, \texttt{camsari@ucsb.edu}, \texttt{tsrimani@andrew.cmu.edu}}
        \affiliation{Department of Electrical and Computer Engineering, Carnegie Mellon University, Pittsburgh, PA 15213, USA}
        
        \date{\today}
        
        \begin{abstract}

    Combinatorial optimization problems are central to science and engineering and specialized hardware from quantum annealers to classical Ising machines are being actively developed to address them. These systems typically sample from a fixed energy landscape defined by the problem Hamiltonian encoding the discrete optimization problem. The recently introduced Probabilistic Approximate Optimization Algorithm (PAOA) takes a different approach: it treats the optimization landscape itself as variational, iteratively learning circuit parameters from samples. Here, we demonstrate PAOA on a 64$\times$64 perimeter-gated single-photon avalanche diode (pgSPAD) array fabricated in 0.35 $\mu$m CMOS, the first realization of the algorithm using intrinsically stochastic nanodevices. Each p-bit exhibits a device-specific, asymmetric (Gompertz-type) activation function due to dark-count variability. Rather than calibrating devices to enforce a uniform symmetric (logistic/tanh) activation, PAOA learns around device variations, absorbing residual activation and other mismatches into the variational parameters. On canonical 26-spin Sherrington-Kirkpatrick instances, PAOA achieves high approximation ratios with $2p$ parameters ($p$ up to 17 layers), and pgSPAD-based inference closely tracks CPU simulations. These results show that variational learning can accommodate the non-idealities inherent to nanoscale devices, suggesting a practical path toward larger-scale, CMOS-compatible probabilistic computers.
        
        \end{abstract}

        \maketitle
        
        
        \section{Introduction}
        Combinatorial optimization problems arise throughout science and engineering, from protein folding to scheduling to circuit design. Over the past decade, specialized hardware has been developed to address these problems, including quantum annealers \cite{johnson2011quantum, king2022coherent}, coherent Ising machines \cite{yamamoto2017coherent, honjo2021100}, and networks of coupled oscillators \cite{wang2019oim, chou2019analog}. More recently, probabilistic computers built from stochastic magnetic tunnel junctions \cite{borders2019integer, kaiser2022hardware}, memristors \cite{pbits_memristors, pbits_rram}, and single-photon avalanche diodes \cite{Whitehead2023NatElec, pbits_SPADs_2} have emerged as CMOS-compatible alternatives that exploit intrinsic device noise for sampling. Despite their diversity, these systems share a common operational principle: they sample from a fixed energy landscape defined by the problem Hamiltonian. The user maps the optimization objective onto coupling weights $\{J, h\}$ \cite{lucas2014ising}, and the hardware explores configurations according to that frozen landscape.
    
        The recently introduced Probabilistic Approximate Optimization Algorithm (PAOA) takes a fundamentally different approach \cite{weitz2025subuniversal, abdelrahman2025probabilistic}.
        Rather than sampling from a fixed Hamiltonian, PAOA treats the optimization landscape itself as variational: circuit parameters are iteratively updated based 
        on sampled configurations to minimize the cost function. Notably, the learned parameters are transferable across instances within the same problem family.  
        This closed-loop framework, inspired by the Quantum Approximate Optimization Algorithm (QAOA) \cite{farhi2014quantum}, uses gradient-free optimization to update couplings based on measured statistics. By adjusting $\{J, h\}$ at each iteration, PAOA reshapes the energy landscape to concentrate probability on low-energy states rather than relying on a static mapping from problem to hardware.
    
      This variational approach addresses two challenges that have limited the scalability of analog probabilistic hardware. The first is device-to-device variability. For example, in an idealized probabilistic circuit (p-circuit), every probabilistic bit (p-bit) would exhibit an identical, symmetric activation function mapping input to output probability \cite{kerem_prx}. Real nanodevices rarely satisfy this assumption. Fabrication and process variations, material inhomogeneities, and environmental fluctuations shift both the slope and operating point of each device's response \cite{MRAM_variations, RRAM_variations, SPAD_variations}. When thousands of p-bits are composed into a network, this activation mismatch distorts the realized distribution and the hardware implements $\{J', h'\}$ rather than the intended $\{J, h\}$. The second challenge is algorithmic: conventional Ising machines and probabilistic computers require hand-tuned annealing schedules and heuristics whose effectiveness depends on problem structure and hardware characteristics \cite{kirkpatrick1983optimization, johnson2011quantum}. Both problems grow more severe as systems scale.

        \begin{figure*}[!t]
            \centering
            \includegraphics[width=\linewidth]{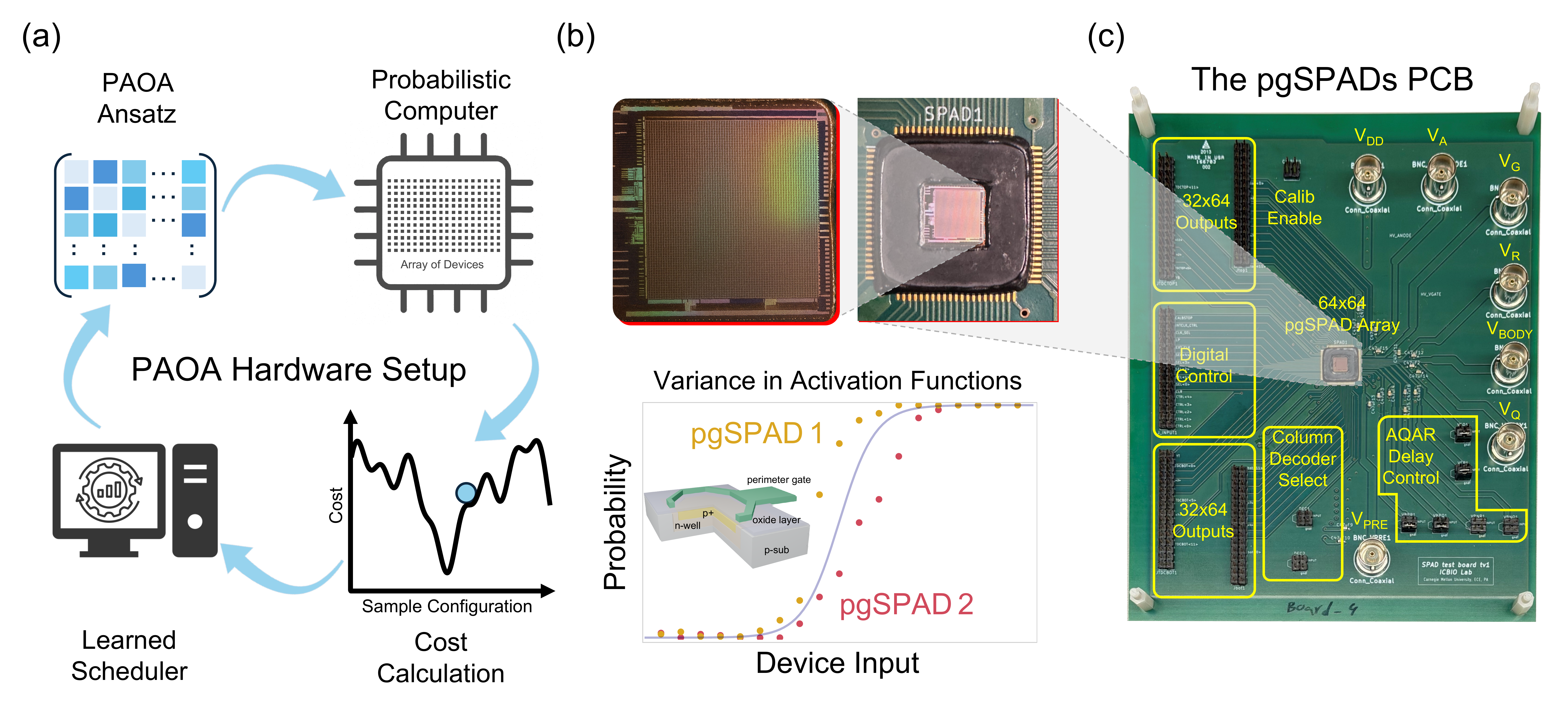}
           
           \caption{\textbf{PAOA implemented on pgSPAD-based probabilistic hardware.} (a) Hardware--software loop: a classical optimizer updates PAOA variational parameters based on a cost computed from sampled configurations. (b) Die micrograph and device-level variability: Different pgSPADs exhibit distinct activation functions (pgSPAD 1 vs pgSPAD 2) due to fabrication-induced variation in dark-count statistics. Inset: cross-section of a perimeter-gated SPAD. (c) Printed circuit board hosting the $64 \times 64$ pgSPAD array with control and readout circuitry.}
           \label{fig:intro}
        \end{figure*}
   
    PAOA addresses both challenges through learning. Because parameters are optimized directly from sampled statistics, the algorithm absorbs device non-idealities into the learned variational parameters. It does not require or assume that p-bits share a common activation function. Instead, it discovers variational parameters (e.g., weights or schedules) that produce the target distribution on the hardware as it actually behaves, not as it is ideally modeled. Similarly, PAOA eliminates the need for manual schedule design: effective annealing strategies emerge from the optimization loop rather than from human intuition. This hardware-aware learning mitigates problems related to device variability and reduces reliance on hand-tuned schedules and heuristics. 
    
    Here, we demonstrate PAOA on a $64 \times 64$ perimeter-gated single-photon avalanche diode (pgSPAD) array fabricated in 0.35~$\mu$m CMOS, the first realization using intrinsically stochastic nanodevices. Each of the 4,096 p-bits exhibits a device-specific, asymmetric (Gompertz-type) activation function arising from dark-count variability, deviating substantially from the ideal logistic response assumed in standard p-bit models as shown in Fig. \ref{fig:intro}. We show that PAOA learns around the asymmetric Gompertz-type activation function, absorbing activation mismatch into the variational parameters without requiring per-device calibration to enforce a uniform (logistic/tanh) response. On canonical 26-spin Sherrington--Kirkpatrick spin-glass instances, PAOA achieves high approximation ratios with $2p$ parameters ($p$ up to 17 layers).
    Notably, inference on pgSPADs closely tracks CPU simulations 
    despite device variations.  
        
    These results establish variational learning as a practical strategy for accommodating the non-idealities inherent to nanoscale devices, providing a path toward scalable, CMOS-compatible probabilistic computers. 
         
        \begin{figure*}[t]
        \centering
        \includegraphics[width=\linewidth,keepaspectratio]{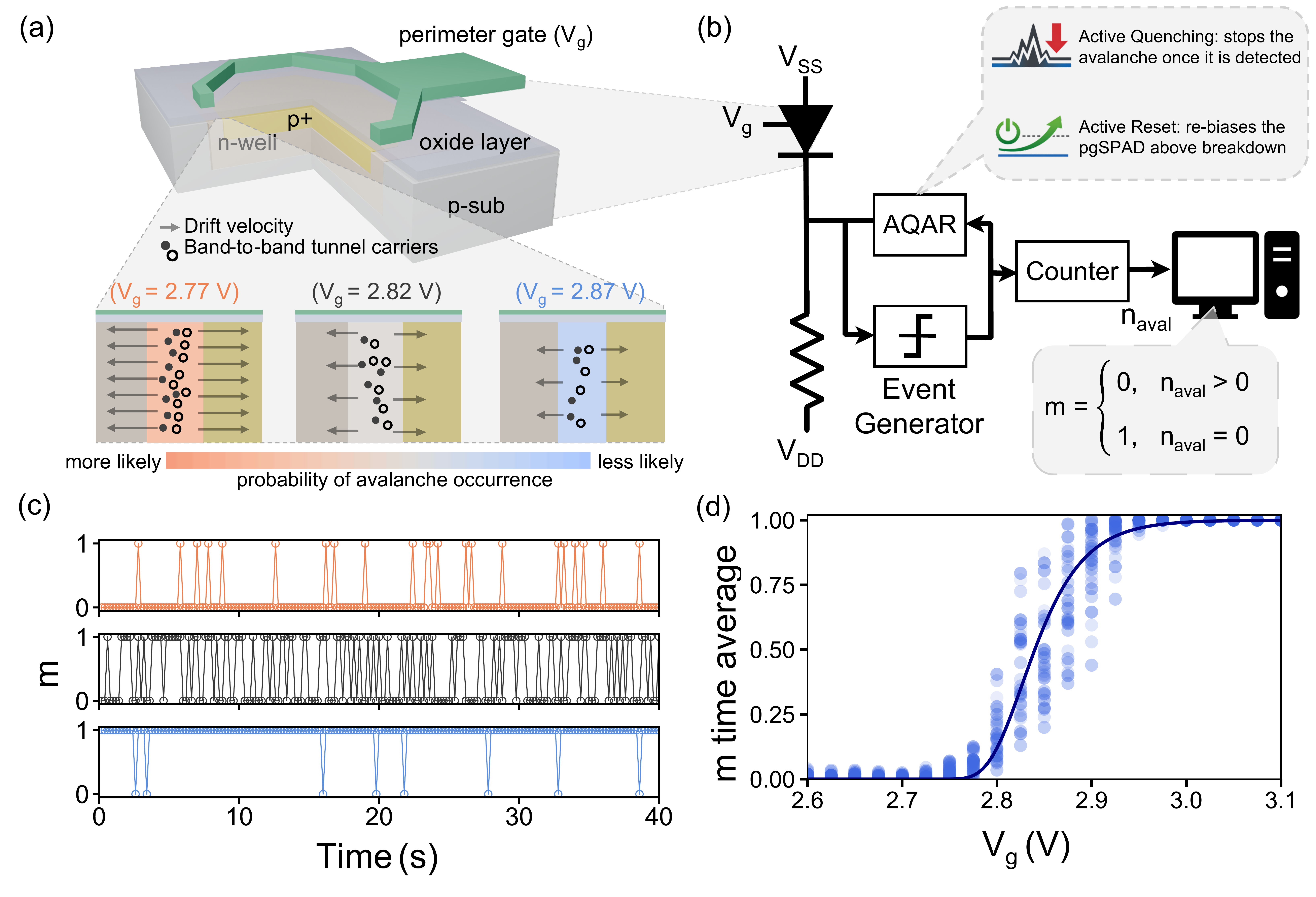}
        \caption{\textbf{Operation and characterization of pgSPAD-based p-bits.} (a) Cross-sectional schematic of a p+/n-well perimeter-gated SPAD. Higher gate voltage $V_g$ reduces the electric field at the junction edge, suppressing dark carrier generation and lowering the avalanche probability. (b) Functional block diagram. The device is biased above breakdown; an active quench-and-reset (AQAR) circuit detects avalanche events and re-arms the device. The binary output $m$ is assigned based on whether at least one avalanche occurs within the integration window. (c) Representative time traces of $m$ for three gate voltages, showing voltage-dependent switching statistics. (d) Measured activation functions for multiple pgSPAD devices, showing device-to-device variability in the sigmoidal response. Solid curve: fitted Gompertz function.}
        \label{fig:variation}
    \end{figure*}

    \begin{figure*}[t]
        \centering
        \includegraphics[width=\textwidth]{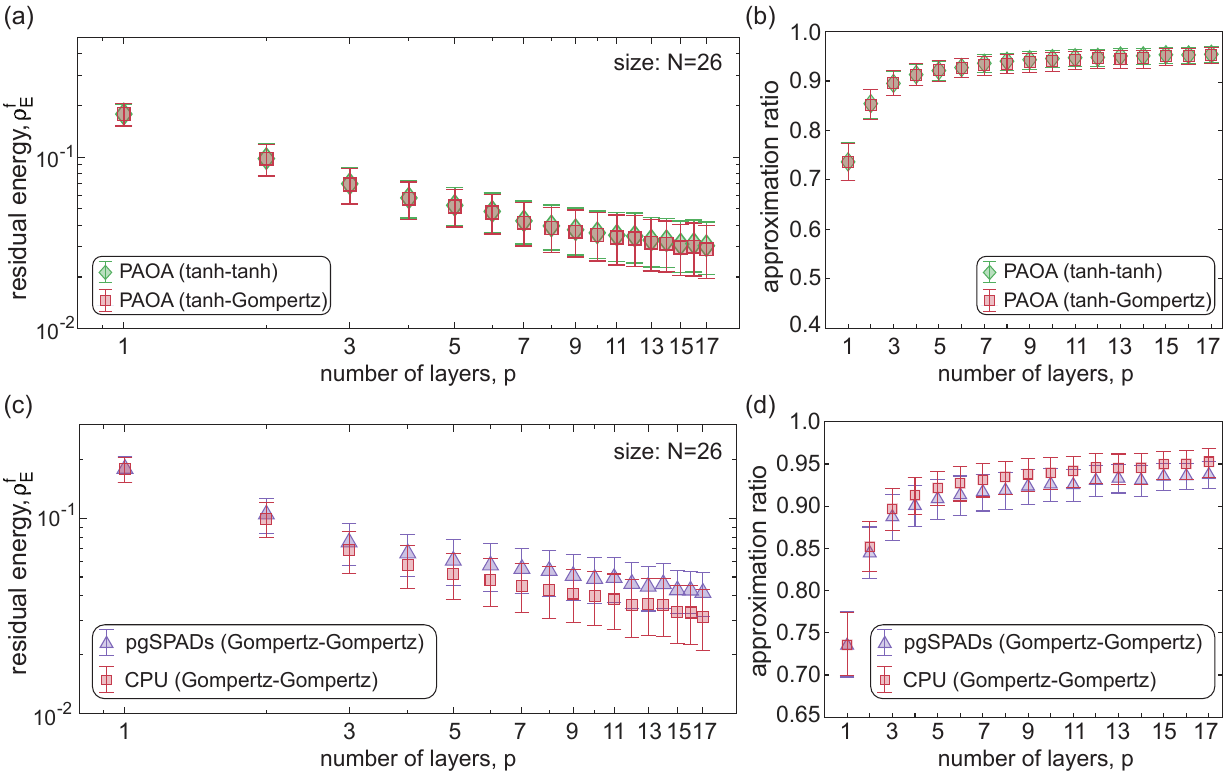}
        \caption{\textbf{PAOA with symmetric and asymmetric activations  benchmarking on 26-spin Sherrington--Kirkpatrick instances.} (a,b) Residual energy and approximation ratio versus depth $p$ for PAOA (tanh–tanh): trained and evaluated with symmetric tanh activation (green), and PAOA (tanh–Gompertz): trained with tanh and evaluated with asymmetric Gompertz activation (red). PAOA schedules are trained on 30 instances and applied without retraining to 30 test instances. Each point averages $10^6$ independent runs. (c,d) Comparison of pgSPAD hardware inference (purple, 50 runs) and CPU simulation (red, $10^6$ runs), both using matched  Gompertz-trained/Gompertz-inferred schedules. Hardware results closely track  simulation despite reduced sampling and device-to-device variability. Error bars: 95\% confidence intervals from bootstrap resampling.}
        \label{fig:exp_paoa}
    \end{figure*}

\section{Perimeter-gated SPADs as p-bits}
        \label{sec:pgspad_pbit}
    
    A perimeter-gated SPAD \cite{Dandin2010,Dandin2016,Dandin2012a}, operated in darkness, functions as a tunable probabilistic bit whose output statistics are controlled by a single gate voltage~\cite{sajaltcas2,pbits_SPADs_2} (Fig.~\ref{fig:variation}). 
    A gate electrode surrounding the junction edge modulates the local electric field and thus the rate of dark carrier generation that can trigger avalanche breakdown (Fig.~\ref{fig:variation}a). Lower gate voltage increases the avalanche probability; higher gate voltage suppresses it. This exponential sensitivity provides a compact physical mechanism for implementing a stochastic activation function.
        
    Each pgSPAD is biased above breakdown and interfaced with an active quench-and-reset (AQAR) circuit that detects avalanche events, quenches them, and re-arms the device for subsequent sampling (Fig.~\ref{fig:variation}b). During a fixed integration window $T_{\mathrm{int}}$, avalanche events are detected and converted into a binary output $m$: we assign $m=0$ if at least one avalanche occurs and $m=1$ otherwise. Repeating this process produces a truly stochastic bitstream whose statistics depend on the gate voltage, as illustrated by representative time traces in Fig.~\ref{fig:variation}c.
        
       Averaging the binary output over many integration windows yields a voltage-controlled activation function. Fig.~\ref{fig:variation}d shows the measured output probability as a function of gate voltage for multiple pgSPAD devices across the array. Each device exhibits a sigmoidal response well described by a Gompertz function (Methods, Eqs.~\ref{eq:lambda_main_text}--\ref{eq:Gompertz_beta}), with noticeable device-to-device variability in slope and offset. The impact of temperature and timing variations on the pgSPAD activation slope and offset is characterized in Supplementary Fig.~\ref{fig:alpha_beta_variation}. Rather than calibrating each device to enforce a standard logistic response, we extract a best-fit hardware activation from experimental samples and allow PAOA to learn directly from this non-ideal response. A controlled study on a four-node majority gate confirms that PAOA learns the correct target distribution under both symmetric (tanh) and asymmetric (Gompertz) activations, with matched-activation training yielding tighter convergence to the correct distribution (Supplementary Section~\ref{PAOA_MAJ_gate}).

    \section{Experiments and Results}
    \label{sec:experiments}
    
    We evaluate PAOA on the Sherrington--Kirkpatrick (SK) spin-glass model, a fully connected random Ising problem that serves as a canonical benchmark for optimization algorithms. The experimental protocol is described in Methods. Note that PAOA can in general variationally parameterize layerwise couplings and biases (see Supplementary Sec.~\ref{sec:MAJ}). For the SK optimization benchmarks here, we use a two-schedule ansatz  in which the variational parameters are inverse-temperature schedules  that effectively rescale the programmed couplings during the anneal.  For the software comparison in Fig.~\ref{fig:exp_paoa}(a,b), schedules  are trained with a symmetric tanh activation. For hardware deployment  in Fig.~\ref{fig:exp_paoa}(c,d), schedules are separately trained using  a uniform Gompertz activation calibrated to the measured array-average response, and then deployed on the pgSPAD array for inference, which  naturally includes device-to-device activation variability. In this work, we train SK schedules in calibrated CPU simulations and deploy the optimized schedules on the pgSPAD array for hardware 
    inference only; the training-inference loop schematic in Fig.~\ref{fig:intro}(a) illustrates the general PAOA workflow.
        
    Following the setup used in \cite{abdelrahman2025probabilistic} for PAOA on the SK problem, we use a two-schedule ansatz with the same $2p$ parameters. We report three PAOA evaluations: (i) tanh–tanh (ideal software baseline), (ii) tanh–Gompertz (activation-mismatch ablation), and (iii) Gompertz–Gompertz, where Gompertz schedules trained in a calibrated CPU model are used for both CPU and pgSPAD hardware inference. We generate 60 random SK instances with i.i.d.\ $J_{ij}\sim\mathcal{N}(0,1)$, compute ground-state energies by exhaustive search, and report residual energy (Eq.~\ref{eq:residual_energy}) and approximation ratio (Eq.~\ref{eq:approximation_ratio}) versus depth $p$.

\begin{figure*}[t!]
        \centering
        \includegraphics[width=\linewidth]{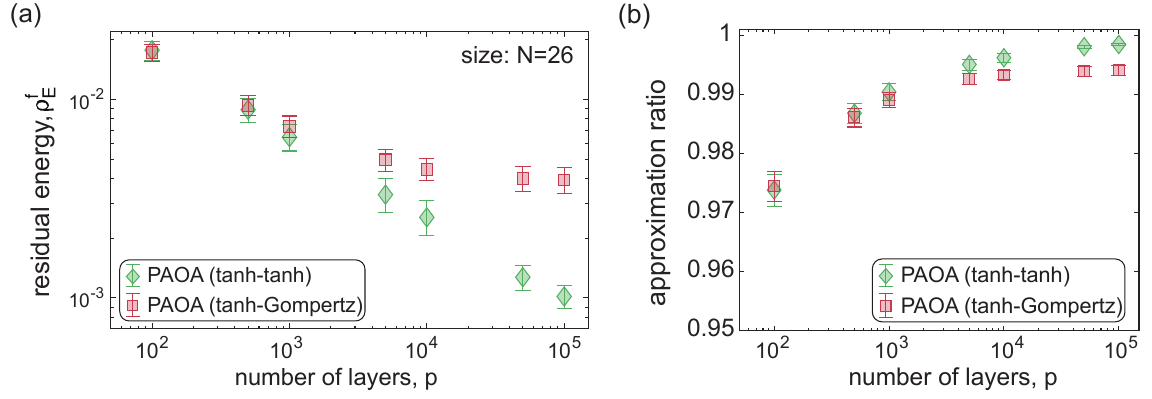}
      \caption{\textbf{Extrapolated PAOA performance at large depth.} (a) Residual energy and (b) approximation ratio versus depth $p \in [10^2, 10^5]$ for PAOA (tanh--tanh) (green) and PAOA (tanh--Gompertz) (red), as defined in Fig.~\ref{fig:exp_paoa}. Schedules learned at $p = 17$ are extrapolated without retraining. Each point averages 300 instances and $10^3$ runs. Error bars: 95\% confidence intervals.}
        \label{fig:extrapolated}
    \end{figure*}

Fig.~\ref{fig:exp_paoa} summarizes the results. Panels (a,b) compare PAOA with matched training and inference using a symmetric tanh activation with mismatched training where inference uses the pgSPAD asymmetric Gompertz activation as a function of depth $p$ on 30 test instances, unused in training. 
Notably, training PAOA with a symmetric activation (tanh) and performing inference with an asymmetric one (Gompertz) produces no discernible performance difference at these depths: the variational schedules are robust to activation mismatch in this regime. The two modes diverge only at much larger depths (Fig.~\ref{fig:extrapolated}), where matched-activation training may become important.
        
 Panels (c,d) in Fig.~\ref{fig:exp_paoa} compare pgSPAD hardware inference against CPU simulation, both using \textit{matched} Gompertz training and inference (i.e., schedules optimized with the calibrated Gompertz activation and then deployed on the Gompertz-native pgSPAD hardware). Despite the large difference in sample count (50 runs per instance on hardware versus $10^6$ on CPU) and the presence of device-to-device variability on the physical array, the hardware results closely track simulation, confirming that real-device non-idealities do not noticeably degrade performance. Although panels~(a,b) in Fig.~\ref{fig:exp_paoa} show that tanh-trained schedules would perform comparably at these depths, using matched training for hardware deployment is the principled choice and becomes increasingly important at larger depths. 
        
For context, rigorous analysis of QAOA on the 
 SK model~\cite{farhi2022qaoa, QAOA_max_cut_farhi} shows that the 
 residual energy decreases with circuit depth as $1/p^a$ with $a=O(1)$. 
 However, the best known method for analytically evaluating the QAOA cost function scales as $O(16^p)$, limiting the depths at which 
 optimal angles have been determined to date. PAOA with $2p$ 
 parameters achieves comparable or lower residual energies at 
 moderate depths~\cite{abdelrahman2025probabilistic}, while each 
 cost function evaluation requires only sampling from the probabilistic 
 circuit.
    
 While Fig.~\ref{fig:exp_paoa} benchmarks PAOA against accessible pgSPAD hardware inference depths, we also examine the asymptotic behavior of PAOA at much larger depths. Fig.~\ref{fig:extrapolated} extends the analysis to depths $p \in [10^2,10^5]$ using CPU simulation with a uniform activation (no per-device variability). Both curves use tanh-trained schedules learned at shallow depth ($p=17$) and extended to deeper depths using a geometric schedule fitting without retraining; the two curves differ only in the activation used during inference (tanh vs.\ Gompertz). More details on the functional form of the fitting are in the Methods section. 
        
Residual energy decreases monotonically with depth, while the approximation ratio approaches unity. 
The tanh--tanh and tanh--Gompertz modes exhibit similar scaling up to approximately $10^3$ layers. At larger depths, the Gompertz inference activation shows mild saturation in both metrics. At moderate depths (Fig.~\ref{fig:exp_paoa}a,b), activation mismatch 
carries no measurable penalty, but at the larger depths probed here 
a gap emerges between tanh and Gompertz inference. Because both  curves share the same tanh-trained schedules, this gap is consistent  with growing sensitivity to activation mismatch rather than a  fundamental limitation of the Gompertz function itself. 

These results demonstrate that PAOA performance is largely insensitive to activation mismatch at moderate depths, and that schedules learned at shallow depth transfer to much deeper regimes. 
pgSPAD-based probabilistic hardware can thus serve as a scalable inference engine for variational optimization, with performance closely tracking CPU simulations.

        \section{Methods}
        \label{sec:methods}
                
        All experiments were performed on a custom $64\times64$ array of \emph{perimeter-gated} single-photon avalanche diodes, operated in the dark and read out with on-board active quench/reset electronics. “Perimeter gating” denotes an independently biased ring that modulates the electric field near the junction perimeter and thus the effective breakdown voltage. Increasing the (magnitude of the) gate bias raises the breakdown threshold and suppresses the dark-count rate (DCR) at fixed reverse bias; decreasing the gate bias has the opposite effect. This knob enables a direct, voltage-programmable mapping from a p-bit’s tuner input to the probability of emitting a logical “1”. pgSPAD architecture, operational timing and the single-event regime used in this work are summarized in Supplementary Secs.~\ref{sec:S_operation}–\ref{sup:variations}.
        
        We bias each pgSPAD above breakdown during the “sense” half-cycle and re-arm it during the “reset” half-cycle using standard active quench/reset timing. Each pixel delivers a binary output $m_i\in\{0,1\}$ once per integration window $T_{\text{int}}$ (CLK period), indicating whether at least one avalanche occurred. The array was controlled through a temperature chamber and multi-channel power supplies; counts were acquired over the full array using a data acquisition circuit (DAQ) and custom software (see Supplementary Fig.~\ref{fig:setup} and Section~\ref{sup:setup} for setup details).

        Our perimeter-gated approach differs from prior SPAD-based p-bits that tune randomness primarily via global bias or front-end quench circuitry. For context on alternative SPAD p-bit implementations and system-level p-computing see for example, Ref.~\cite{Whitehead2023NatElec,whitehead2024cmos}. This distinction is important because perimeter gating provides a device-local, exponential control of dark generation \emph{without} changing the global reverse bias, which we exploit for per-pixel calibration and PAOA control (Supplementary Section~S.3).
        
        \subsection{Activation function and per-pixel calibration}
        
        The probability that pixel $i$ emits $m_i{=}1$ in a window $T_{\text{int}}$ is determined by its dark-count process. Empirically, the DCR obeys
        \begin{equation}
        \lambda_i(\theta,V_g)\;=\;\lambda_{0,i}\,\exp\!\big(\zeta_i\,\theta-\alpha_i\,V_g\big),
        \label{eq:lambda_main_text}
        \end{equation}
        with $\theta$ temperature, $V_g$ the perimeter-gate bias, and $\alpha_i,\zeta_i$ device-specific coefficients. Here, $\lambda_{0,i}$ is the native DCR at $0^oC$ for a $0~V$ on the gate. Assuming Poisson statistics for counts within $T_{\text{int}}$, the probability of one-or-more events is
        \begin{equation}
        P_{DC,i}\;=\;1-\exp\!\big(-\lambda_i(\theta,V_g)\,T_{\text{int}}\big),
        \label{eq:PDC_text}
        \end{equation}
        and the p-bit activation (probability of logical “1”) becomes
        \begin{equation}
        \Pr\{m_i{=}1\}\;=\;1-P_{DC,i}\;=\;\exp\!\Big(-\kappa_i\,e^{-\alpha_i V_g}\Big)
        \label{eq:Gompertz_text}
        \end{equation}
        
        \vspace{-10 pt}
        
        \begin{equation}
        \kappa_i=\lambda_{\theta,i}T_{\text{int}}; ~~ \lambda_{\theta,i} = \lambda_{0,i}e^{\zeta_i\theta},
        \label{eq:Gompertz_beta}
        \end{equation}
        a Gompertz-type sigmoid. Equation~(\ref{eq:Gompertz_text}) is the device-level activation implemented by the pgSPAD p-bits and used to calibrate the sampler for PAOA training and hardware inference.

         We use three convenient descriptors throughout: (i) the mid-point $V_{\text{mid}}$ at $\Pr\{m_i{=}1\}{=}0.5$, (ii) the inflection $V_{\text{inflc}}$ where $d^2\Pr/dV_g^2{=}0$, and (iii) the transition range $\Delta V_{\text{transit}}$ defined by the extrema of $d^2\Pr/dV_g^2$. Closed forms and derivations are provided in Supplementary Section~\ref{sup:equations} (Eqs.~S.1–S.6), including the linear ($\theta$) and logarithmic ($T_{\text{int}}$) shifts of operating points.

        Prior to learning, each pixel is independently swept in $V_g$ at the chosen $T_{\text{int}}$ and temperature. We fit $\Pr\{m_i{=}1\}(V_g)$ to extract $(\alpha_i,\kappa_i)$ and compute $(V_{\text{mid},i},V_{\text{inflc},i},\Delta V_{\text{transit},i})$. During operation we center the pixel at $V_{\text{bias},i}{=}V_{\text{mid},i}$ and drive small-signal excursions about this point. Temperature drift $\delta\theta$ induces a predictable shift $\delta V_{\text{mid},i}\!\approx\!(\zeta_i/\alpha_i)\,\delta\theta$, which we compensate in software by updating $V_{\text{bias},i}$; jitter and drift analyses appear in Supplementary Section~\ref{sup:variations}.
        
        The array is partitioned into independent cohorts that update and are read out in parallel. A cohort is defined by (i) a set of pixels that realize a single p-circuit instance and (ii) a schedule (clocking and bias updates) that is globally applied but locally parameterized by $\{V_{\text{bias},i},k_i\}$. 
        
        This per-pixel centering is a hardware-level setup step independent  of PAOA; the algorithm itself trains on a single array-average  Gompertz activation and does not require or use per-device  activation parameters. One PAOA “experiment” consists of programming the weights $(J,h)$ into all cohorts, collecting $N$ samples from each cohort over $N$ integration windows, forming histograms of the observed states, and returning empirical probabilities $p_m$ to the optimizer. For distribution-learning tasks (e.g., majority gate), we use a cross-entropy cost with  Laplace smoothing ($\epsilon$ added to each count), whereas we minimize the average energy for optimization.

        \subsection{Mapping p-circuit inputs to pgSPAD gate voltages}
        
        We implement the p-circuit update by modulating each pixel’s gate around its bias:
        \begin{equation}
        V_{g,i}\;=\;V_{\text{bias},i}\;+\;k_i\,I_i,
        \label{eq:VG_map_text}
        \end{equation}
        where the tuner input $I_i$ is
        \begin{equation}
        I_i\;=\;\sum_{j}J_{ij}\,s_j\;+\;h_i.
        \label{eq:I_text}
        \end{equation}
        Here $J_{ij}$ and $h_i$ are the programmed couplings and biases (for the SK instances studied here, $h_i$=0). The scale factor $k_i$ converts the dimensionless tuner input to volts. We choose $k_i$ so that the \emph{local} slope of the Gompertz activation at the operating point matches that of a logistic/tanh p-bit; since the Gompertz slope at $V_{\text{inflc},i}$ equals $\alpha_i/e$ and the tanh-based logistic slope at its inflection is $1/2$ with respect to its input, matching slopes yields
        \begin{equation}
        k_i\;=\;\frac{e}{2\,\alpha_i}.
        \label{eq:slope_match_text}
        \end{equation}
        In practice we set $V_{\text{bias},i}=V_{\text{mid},i}$ and use the same $k_i$ (or a small neighborhood-dependent adjustment) across the experiment; this linearization is sufficient for the small excursions used by PAOA. Additional discussion of the activation function asymmetry and a symmetric surrogate used in software ablations is provided in Supplementary Section~S.2.
    
        The network samples configurations weighted by the Ising energy
    \begin{equation}
    E(\{s\}) = -\frac{1}{\sqrt{N}}\sum_{i<j} J_{ij} s_i s_j ,
    \label{eq:ising_energy}
    \end{equation}
    where $s_i \in \{-1,+1\}$ are Ising spins related to binary hardware states by $s_i = 2m_i - 1$. For a given instance, we minimize the J-averaged measured energy $\langle E(\{s\}) \rangle_J$, where $E(\{s\})$ is the Ising energy of state $s$ calculated using the instance weight matrix $J$.
        
    \subsection{PAOA training and extrapolation}
        PAOA training consists of inner and outer loops. In the inner loop (i.e., at each optimizer iteration), PAOA runs for the desired $p$ layers and collects final samples at the $p^{\text{th}}$ layer from independent runs. In the outer loop, a gradient-free optimizer, Constraint Optimization by Linear Approximation (COBYLA), updates the variational parameters iteratively \cite{powell1994cobyla, arrasmith2021effect, schiffer2022adiabatic}. COBYLA perturbs the variational parameters in the direction that reduces the average energy $\langle E(\{s\})\rangle$. The stopping criteria is reached when either the perturbation step-size falls below a predefined threshold $\epsilon_{\text{step}}$ or a maximum iteration budget is exhausted. The full training pseudocode is outlined in Algorithm~\ref{algo:PAOA}. The hyperparameter values used in the training are listed in Table.~\ref{tab1:simulation_parameters}.  See Supplementary Section \ref{sup:algo} for further details.
        
        For the PAOA results in Section~\ref{sec:experiments}, we follow the same method as in \cite{abdelrahman2025probabilistic}. We generate 60 random SK all-to-all instances of size $N=26$ with couplings $J_{ij}\sim \mathcal{N}(0,1)$ sampled from a standard normal distribution. 
    
        The two-schedule PAOA ansatz is trained separately for each instance. For inference, we use the schedule obtained by averaging the learned schedules across the 30 training instances and apply it to 30 freshly generated test instances. Finally, we report the residual energy $\rho^f_E$ defined as a function of PAOA depth $p$:
        \begin{equation}
            \rho^f_E(p) = \frac{\langle E(p)- E_{\text{sol}}\rangle_J }{N},
            \label{eq:residual_energy}
        \end{equation}
        and approximation ratio defined as: 
        \begin{equation}
            \text{Approx.  ratio} = \frac{E(p)}{E_{\text{sol}}},
            \label{eq:approximation_ratio}
        \end{equation}
        where $E_{\text{sol}}$ denotes the ground-state energy obtained by exhaustive enumeration. Error bars indicate $95\%$ confidence intervals computed via bootstrapping. 
    
        For deeper PAOA depths, we extrapolate the schedules optimized at shallow depth ($p=17$) using a geometric schedule with the following functional form:
       \begin{equation}
            \log \beta^{\text{model}}_k = \log(\beta_0) + \gamma_k\log\left(\frac{\beta_f}{\beta_0}\right)\
            + c\,\gamma_k(1-\gamma_k),
        \end{equation}
        where $\beta_0$ and $\beta_f$ denote the initial and final values of the fitted schedule,
        respectively, and $\gamma_k = k/(p-1)$ for $k=0,\dots,p-1$.
        The parameter $c$ is determined by least-squares fitting to the optimized
        schedule at $p=17$, by minimizing the squared error,
        \begin{equation}
            c^* = \underset{c}{\text{argmin}} \|\log \beta^{\mathrm{opt}} - \log \beta^{\text{model}}\|^2.
        \end{equation}
        This yields the closed-form solution
        \begin{equation}
            c^* =
            \frac{ \bigl(\gamma(1-\gamma)\bigr)^{\!\top}
            \left[\log \beta^{\mathrm{opt}} - \log(\beta_0) - \gamma\log\left(\displaystyle \frac{\beta_f}{\beta_0}\right)\right]}
            { \bigl(\gamma(1-\gamma)\bigr)^{\!\top}\bigl(\gamma(1-\gamma)\bigr)}.
        \end{equation}
        The resulting model is then used to generate schedules for arbitrary PAOA depths
        $p$ using $\gamma_k$.

        \section*{Data and Code availability}
        All data and code used to generate the results and plots in this study are openly accessible at:
        \url{https://github.com/tathagatasrimani/PAOA_SPADs}
        
        \begin{acknowledgments}
        This work was supported in part by National Science Foundation Grant No. 2442346. We further acknowledge support from Samsung and NSF FuSe2 2425218. KYC, ASA and SC acknowledge National Science Foundation (NSF) under award number 2311295, and the Office of Naval Research (ONR), Multidisciplinary University Research Initiative (MURI) under Grant No. N000142312708. FM was partially supported by the Office of Naval Research (ONR) under Award No. N00014-23-1-2771. Use was made of computational facilities purchased with funds from the National Science Foundation (CNS-1725797) and administered by the Center for Scientific Computing (CSC). The CSC is supported by the California NanoSystems Institute and the Materials Research Science and Engineering Center (MRSEC; NSF DMR 2308708) at UC Santa Barbara.
        \end{acknowledgments}
        
        \section*{Author contributions}
        TS, KYC and MD conceived the study. ZA and ASA implemented the PAOA algorithm for the pgSPAD chip. MD designed the pgSPAD chip, and KCL, MSS, and HG designed the boards. MSS and HG conducted pgSPAD characterization under various bias and temperature conditions. ZA conducted all experimental measurements for PAOA implementation on pg-SPADs with input from MSS and HG. ASA performed the PAOA analysis with input from SC. All authors contributed to analyzing the results and writing the manuscript. ZA, ASA, and MSS are co-first authors.
        \vspace{-5pt}
        \section*{Competing interests}
        The authors declare no competing interests. \vspace{-15pt}
        
        \bibliographystyle{apsrev4-2}

        \onecolumngrid
        
        \begin{center}
          \vspace*{2ex}
          {\sffamily\Large\bf Supplementary Information\par}
          \vspace{0.5ex}
          {\sffamily\large\bf Probabilistic approximate optimization using single-photon avalanche diode arrays\par}
          \vspace{1ex}
          {\sffamily Ziyad Alsawidan, Abdelrahman S. Abdelrahman, Md Sakibur Sajal, Shuvro Chowdhury, Kai-Chun Lin, Hunter
    Guthrie, Sanjay Seshan,  Shawn Blanton, Flaviano Morone, Marc Dandin, Kerem Y. Camsari, and Tathagata Srimani}
        \end{center}
        
        \setcounter{section}{0}
        \setcounter{subsection}{0}
        \setcounter{figure}{0}
        \setcounter{table}{0}
        \setcounter{equation}{0}
        
        \renewcommand{\thesection}{S.\arabic{section}}
        \renewcommand{\thesubsection}{S.\arabic{section}.\arabic{subsection}}
        \renewcommand{\theequation}{S.\arabic{equation}}
        \renewcommand{\thetable}{S\arabic{table}}
        \renewcommand{\thefigure}{S\arabic{figure}}
        \renewcommand\figurename{Supplementary Fig.}
                \makeatletter
        \renewcommand{\fnum@figure}{\textbf{Supplementary Fig.~\thefigure}}
        \makeatother

        \setlength{\textfloatsep}{8pt plus 2pt minus 2pt}
        \setlength{\intextsep}{8pt plus 2pt minus 2pt}

        \section{pgSPAD p-bit: architecture and operation}
        \label{sec:S_operation}
        
        A single photon avalanche diode (SPAD) is a highly sensitive p-n device that is capable of detecting individual photons by operating in the "Geiger mode". In this mode, its reverse bias voltage is set above the nominal breakdown voltage. When a single photon is absorbed, it generates an electron-hole pair, also known as the photo-generated carriers that trigger a self-sustaining avalanche. The avalanche of carriers creates a large current that can be easily detected by the front-end electronics, and the event can be registered with a very high timing resolution, \textit{i.e.}, on the order of picoseconds. 
        
        After detection, the diode must be quenched to stop the avalanche and reset for the next detection. Within a specified active time window, the device can be actively quenched and actively reset to continue accumulating the number of detections. However, the device can also be armed for only one detection until a manual reset is performed at the very end.
        
        Aside from the photo-generated carriers, there could be avalanche triggering carriers uncorrelated to any photon incident. These are known as the dark carriers which govern the noise characteristics of these devices. In planar devices, the probability of avalanche triggering by a dark carrier is promoted by a phenomenon known as premature edge breakdown (PEB). The non-uniform electric field in a planar SPAD is the strongest at the curved edges or perimeter of the junction. This causes the perimeter to reach its breakdown voltage before the bulk junction. This PEB effect has several negative consequences, including increased dark count rate (DCR) and lower photon detection efficiency (PDE). Both of these lead to lower sensitivity and higher noise.

        \subsection{Typical SPAD vs pgSPAD}

\begin{figure}[h]
            \centering
            \includegraphics[width=0.8\linewidth]{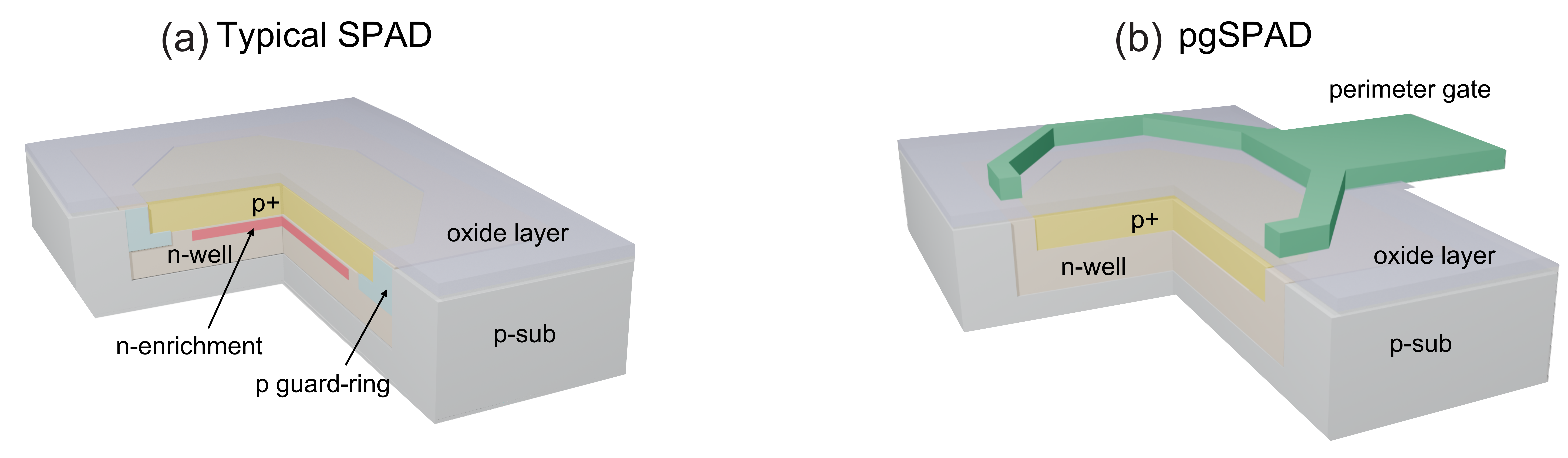}
            \caption{Cross-sectional view of a typical SPAD (a) and a pgSPAD (b).}
            \vspace{10pt}
            \label{fig:spad_compare}
        \end{figure}

        Fig.~\ref{fig:spad_compare} (a) and (b) shows the cross-sectional view of a typical SPAD and a perimeter gated SPAD (pgSPAD), respectively, using p+/n-well configuration. The optional n-enrichment layer (see Supplementary Fig.~\ref{fig:spad_compare} (a)) is used as the multiplication region to promote the avalanche build-up process. To reduce the PEB effect, guard-rings are constructed around the perimeter junction to taper the local electric field. In some processes, shallow trench isolation (STI) is used to achieve the same goal. However, both of these processes are technology dependent. 
        
        The pgSPAD shown in Fig.~\ref{fig:spad_compare} (b) uses a technology-agnostic approach to combat PEB by deploying a polysilicon gate over the perimeter junction. A typical pgSPAD features this polysilicon gate over the oxide layer at the perimeter to taper the electric field underneath when energized. For a p+/n-well SPAD, this gate straddles the perimeter of the p+ anode. As for the shape, these devices are often designed with a circular or octagonal active area, as simulations show these shapes provide a more uniform electric field distribution at the junction compared to rectangular designs. 
        
        By applying a negative bias to the gate relative to the cathode, a depletion region is created at the corner of the junction. This reduces the electric field's strength at the edges, forcing the entire avalanche region to break down more uniformly. As such, this gate terminal can be used as a user input to actively modulate the noise characteristics of the pgSPAD devices. The ability of noise modulation depends on the applied reverse bias voltage and the operating temperature as explained in the next subsection.

        \subsection{Dark noise modulation mechanism in pgSPADs}
        
        \begin{figure}[h]
            \centering
            \includegraphics[width=0.95\linewidth]{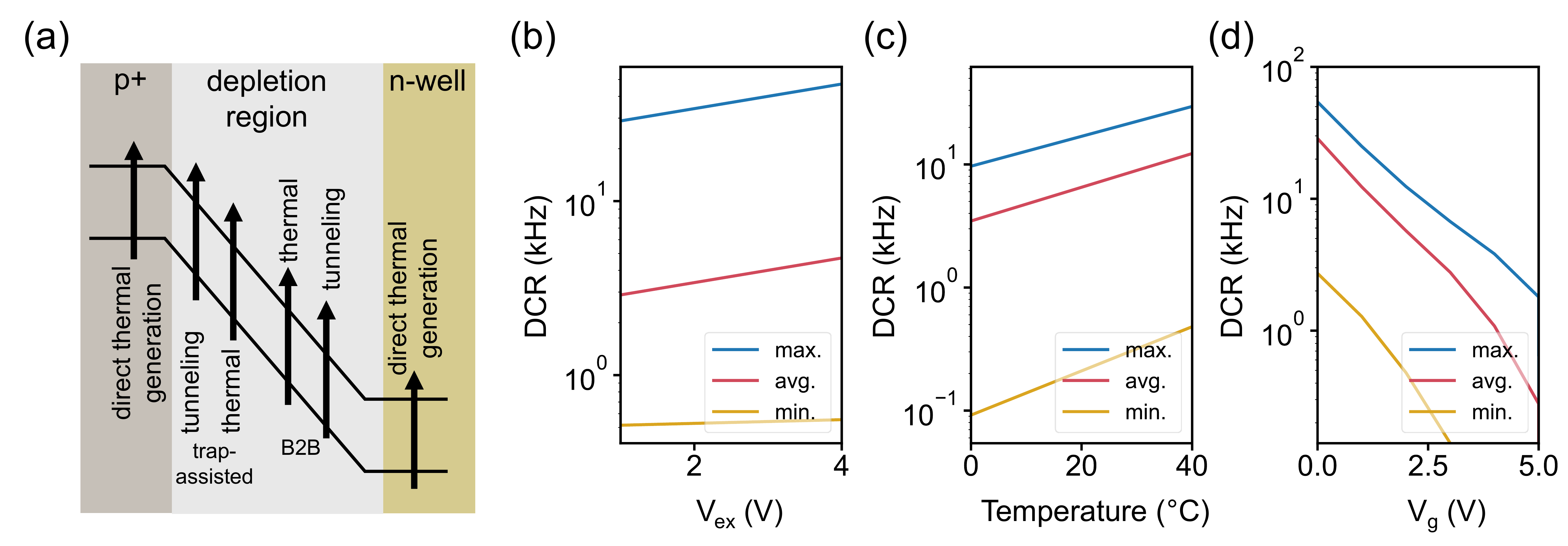}
            \caption{(a) Dark carrier generation mechanism in SPAD devices, and (b-d) dark count rate (DCR) with respect to the excess bias voltage ($V_{ex}$), temperature and the gate voltage ($V_g$) from a pgSPAD array showing the maximum, minimum and the average DCR.}
            \vspace{10pt}
            \label{fig:dark_mecha}
        \end{figure}
        
        \begin{figure}[b]
            \centering
            \includegraphics[width=0.95\linewidth]{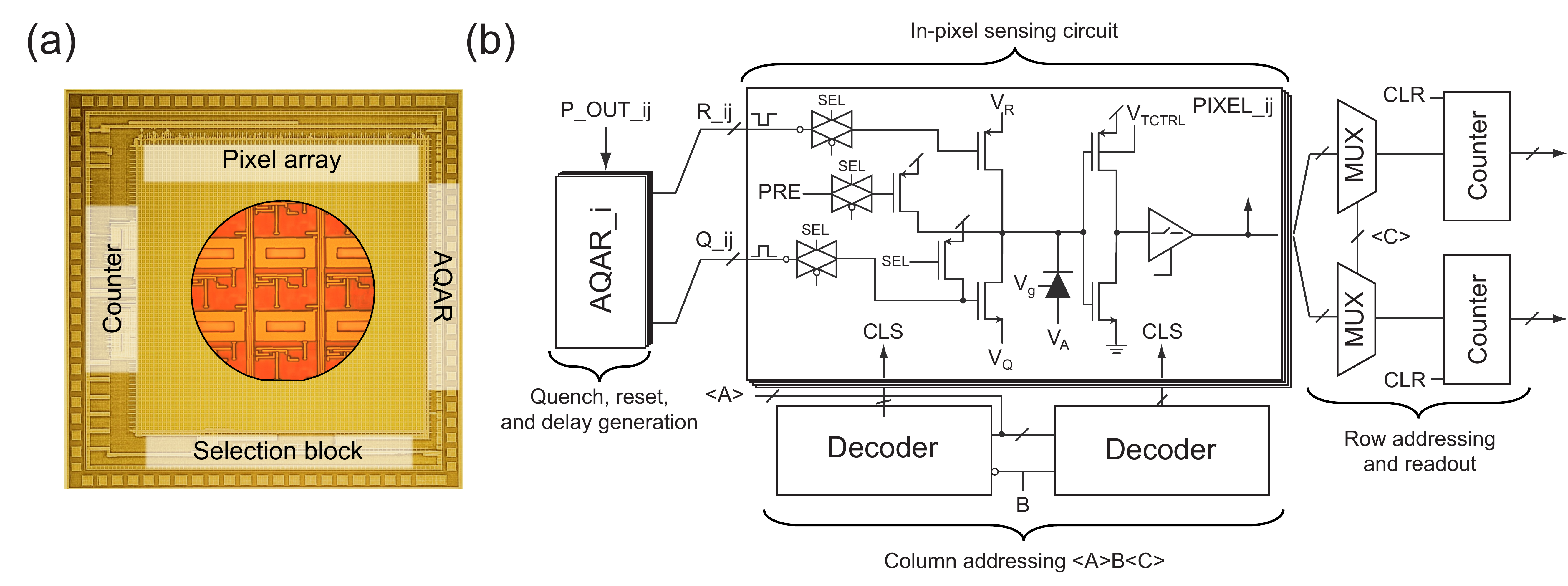}
            \caption{(a) Chip photomicrograph and (b) pixel circuit schematic with the peripheral functional blocks.}
            \label{fig:micrograph}
            \vspace{10pt}
        \end{figure}
        
        Fig.~\ref{fig:dark_mecha} (a) shows the dark carrier generation mechanisms in a SPAD device. These mechanisms can be categorized in thermal generation and band-to-band (B2B) tunneling (direct or trap-assisted). The dominating mechanism depends on the operating condition. For example, at a lower temperature and a higher excess bias voltage, the DCR is dominated by B2B tunneling. However, at higher temperatures, excessive thermal generation dominates the total dark carrier generation rate. 
        
        At high electric fields, carriers can easily tunnel directly from the valence band to the conduction band without requiring thermal energy. Hence, the probability of B2B tunneling is extremely sensitive to the electric field strength in the depletion region. As the excess bias voltage increases, the electric field grows stronger, dramatically increasing the likelihood of a carrier tunneling and initiating an avalanche. Hence, we see an exponential growth of the array average, maximum and minimum of the DCR are shown in Fig.~\ref{fig:dark_mecha} (b) from a randomly selected pgSPAD array containing $4,096$ devices. In modern, deep-submicron CMOS SPADs, where the junctions are often narrow, B2B tunneling is sometimes the dominant source of the DCR.
        
        Similarly, temperature increases the DCR in SPADs primarily by increasing the rate of thermally generated carriers within the device's junction. Hence, these thermally generated carriers, particularly those created through Shockley-Read-Hall (SRH) recombination, can continue to trigger avalanches even without incident photons. As the temperature rises, the rate of this thermal generation, which follows a Boltzmann-like distribution, increases exponentially, leading to a higher frequency of dark counts as seen in Fig.~\ref{fig:dark_mecha} (c). This can indirectly promote trap-assisted B2B tunneling as well. However, by energizing the perimeter gate, it is possible to lower the DCR by effectively reducing the electric field's contribution at the perimeter junction as seen in  Fig.~\ref{fig:dark_mecha} (d).

        \subsection{pgSPAD sensing circuit with active quench and active reset front end}
        
        A pgSPAD array's die photomicrograph is shown in Fig.~\ref{fig:micrograph} (a). The array includes 4,096 p+/n-well pgSPADs are tiled in a 64 $\times$ 64 grid. Each pixel is $\sim$50 $\mu m$ $\times$ 50 $\mu m$ and is individually-accessible for data collection by the row and the column selection block. At most two pixels can be read out simultaneously using the readout counters. Active quenching and active reset are achieved dynamically by routing the pixel's output signal to a row-wise active-quenching active-reset (AQAR) circuit. 
        
        The imager's architecture is shown in Fig.~\ref{fig:micrograph} (b). During operation, a specific pixel address \textbf{$<A>B<C>$} is asserted followed by a clear signal (CLR) which resets the output counters to zero. A hard reset signal (PRE) is then asserted to pre-charge the cathodes of the selected pgSPADs to $V_{DD}$. Upon the pre-charge signal being de-asserted, the pgSPADs actively quench and reset based on avalanche events occurring within their junctions. The output counters are monitored using a data acquisition board.
         
        \section{Methods details: device equations and timing}
        \label{sup:equations}
        
        For a given excess bias voltage, the DCR of pixel $i$ at temperature $\theta$ and perimeter gate voltage $V_g$ is modeled as
        \begin{equation}
        \lambda_i(\theta,V_g)=\lambda_{0,i}\,e^{\zeta_i\theta-\alpha_i V_g}.
        \label{eq:S_dcr}
        \end{equation}
        Here, $\lambda_{0,i}$ is the DCR at $0^oC$ and $0~V$ on the gate while $\zeta_i$ and $\alpha_i$ are the temperature and the gate exponent coefficients, respectively. The rationale for modeling DCR in this way follows from the exponential increase and decrease of it as a function of temperature and gate voltage magnitude shown in Fig.~\ref{fig:dark_mecha} (c-d).
        
        Since dark noise follows a Poisson statistics, within an active window $T_{\mathrm{int}}$ the probability of observing at least one event is
        \begin{equation}
        P_{\mathrm{DC},i}=1-\exp\!\big(-\lambda_i(\theta,V_g)\,T_{\mathrm{int}}\big).
        \label{eq:S_pdc}
        \end{equation}
        
        \subsection{From DCR to p-bit activation (derivations)}
        
        \begin{figure}[b]
            \centering
            \includegraphics[width=\linewidth]{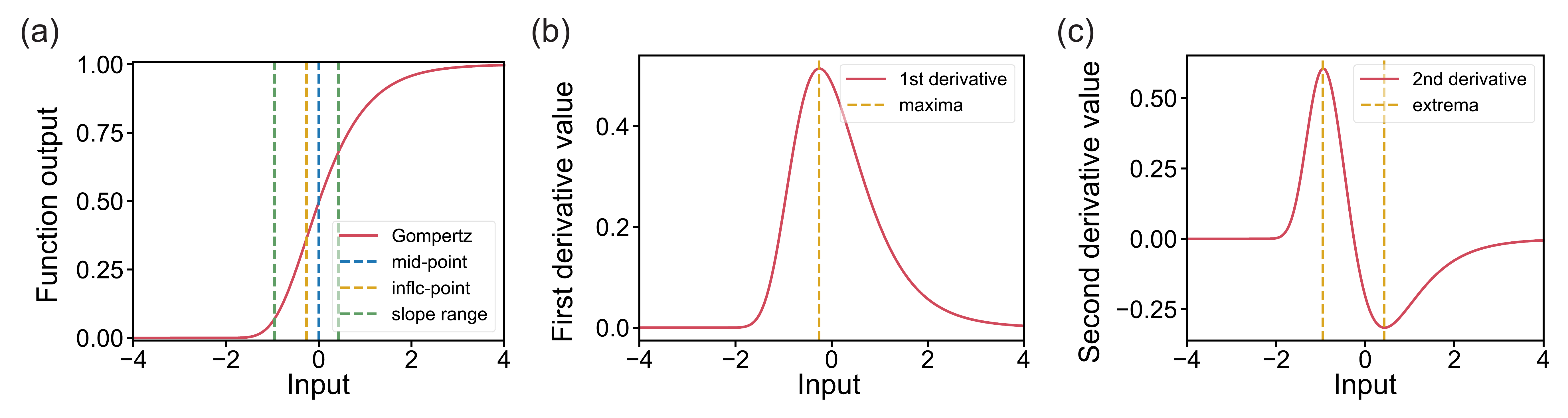}
            \caption{(a) Gompertz activation $p(V)$; (b) first derivative; (c) second derivative. The mid-point $V_{\mathrm{mid}}$, inflection $V_{\mathrm{inflc}}$, and transition range $\Delta V_{\mathrm{transit}}$ are indicated.}
            \label{fig:derivation}
            \vspace{10pt}
        \end{figure}
        
        As the p-bit output defined to be $m_i\in\{0,1\}$, the activation function becomes:

        \begin{equation}
            Pr\{m_i = 1\} = 1 - P_{DC,i} = \exp(-\kappa_ie^{-\alpha_i V_g}), \qquad \kappa_i \equiv \lambda_{\theta,i}T_{int};~ \lambda_{\theta,i}=\lambda_{0,i}e^{\zeta_i\theta}.
            \label{eq:activation_main}
        \end{equation}
        
        \textit{i.e.}, a Gompertz-type sigmoid as shown in Fig.~\ref{fig:derivation} 
        (a). Three useful descriptors follow in closed form:
        \begin{align}
        V_{\mathrm{inflc},i}&=\frac{\ln \kappa_i}{\alpha_i} 
        \label{eq:S_vinfl}\\
        V_{\mathrm{mid},i}&=\frac{\ln(k\kappa_i)}{\alpha_i},\quad k=\frac{1}{\ln 2} 
        \label{eq:S_vmid}\\
        \Delta V_{\mathrm{transit},i}&=\frac{1.925}{\alpha_i}
        \label{eq:S_vwidth}
        \end{align}
        \newline
        \newline
        
        Below we collect the intermediate steps for the derivations of these descriptors.
        
        \paragraph{Inflection Point:}
        Let $p(V)\equiv \exp(-\kappa e^{-\alpha V})$ with the device index omitted for brevity. Then the first derivative is
        \begin{equation}
        \frac{dp}{dV}=\alpha\kappa\,e^{-\alpha V}\,p.
        \label{eq:S_dp}
        \end{equation}
        The inflection satisfies $\frac{d^2p}{dV^2}=0$ as seen in Fig.~\ref{fig:derivation} 
        (b), which yields $\kappa e^{-\alpha V}=1$ and gives Eq.~\eqref{eq:S_vinfl}.
        
        \paragraph{Mid-Point:} The mid-point $V_{\mathrm{mid}}$ solves $p(V_{\mathrm{mid}})=1/2$, i.e.,
        $e^{-\kappa e^{-\alpha V_{\mathrm{mid}}}}=\tfrac{1}{2}$, which leads to Eq.~\eqref{eq:S_vmid}.
        Unlike the logistic function, the mid-point and the inflection point of the Gompertz function do not coincide, making the activation function asymmetric. 
        
        \paragraph{Transition Input Range:} The transition range $\Delta V_{\mathrm{transit}}$ is defined as the distance between the two $V$ values required to flip the pinned output.; solving the extrema seen in Fig.~\ref{fig:derivation} 
        (c) of the second derivative gives the compact approximation in Eq.~\eqref{eq:S_vwidth} (accurate to within a few percent over the range of $\kappa$ encountered here).
        
        \begin{figure}[b]
            \centering
            \includegraphics[width=\linewidth]{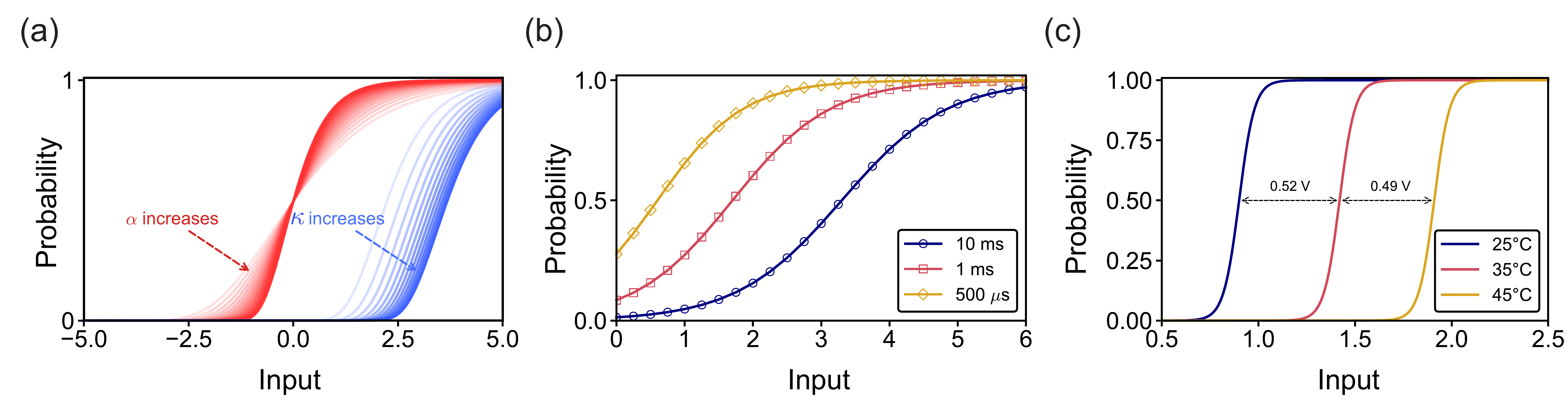}
            \caption{Impact of manufacturing variability on pgSPAD activation characteristics.
            Manufacturing-induced device variations lead to changes in the effective activation function parameters $\alpha$ and $\kappa$ across different pgSPAD units.
            (a) Simulated activation functions illustrating how variations in $\alpha$ primarily affect the slope of the transition region, while variations in $\kappa$ shift the midpoint of the probabilistic response.
            (b) Measured dark-count-induced switching probability as a function of input for different sampling windows, showing how integration time modifies the effective activation curve.
            (c) Measured effect of temperature variation on the activation function, demonstrating shifts in the midpoint and transition range of the probability response as temperature increases.}
            \label{fig:alpha_beta_variation}
            \vspace{10pt}
        \end{figure}
        
        \section{Factors affecting $(\alpha,\kappa)$ and the operating points}
        \label{sup:variations}
        
        \paragraph{Stemming from manufacturing variations:}
        Fig.~\ref{fig:alpha_beta_variation} (a) illustrates representative activation probability curves as a function of the input, highlighting the separate roles of $\alpha$ and $\kappa$. As $\alpha$ increases, the transition becomes steeper, whereas increasing $\kappa$ shifts the curve horizontally. Thus, variations in $\alpha$ and $\kappa$ lead to different transition slopes and mid-points, respectively, resulting in device-to-device variability in the activation probability.
        
        Interestingly, the p-bits can incorporate individual offset bias control to calibrate the $V_{g_{\mathrm{mid}}}$’s so that the tuner input can be assumed to be a small signal around them. This effectively recenters curves with different $\kappa$ values to a common mid-point. The remaining effect of $\alpha$ variation manifests as differences in the transition input range when compared to the ideal logistic function.
        
        \paragraph{Stemming from operating condition fluctuations:}
        
        From Eqs.~\eqref{eq:activation_main} and \eqref{eq:S_vinfl} with $\kappa=\lambda_0 e^{\zeta \theta}T_{\mathrm{int}}$, a change in the integration time $T_{\mathrm{int}}$ shifts the inflection point (as well as the mid-point), consistent with the horizontal displacements observed in Fig.~\ref{fig:alpha_beta_variation}(b). A small timing error (jitter) $\delta T$ modifies the inflection voltage as
        \begin{equation}
        V_{\mathrm{inflc}}(T_{\mathrm{int}}+\delta T) = \frac{\ln(\lambda_\theta[T_{\mathrm{int}} + \delta T])}{\alpha} \approx V_{\mathrm{inflc}}(T_{\mathrm{int}}) + \frac{1}{\alpha}\frac{\delta T}{T_{\mathrm{int}}}.
        \label{eq:S_jitter_shift}
        \end{equation}
        Here we have utilized the series expansion of the logarithm assuming $\frac{\delta T}{T_{\mathrm{int}}} \ll 1$, which is practically reasonable.
        
        Similarly, a fluctuation $\delta\theta$ in the operating temperature alters $\kappa$ and consequently shifts the inflection point (as well as the mid-point), as illustrated in Fig.~\ref{fig:alpha_beta_variation}(c), where increasing temperature leads to a systematic probability shift:
        \begin{equation}
        V_{\mathrm{inflc}}(\theta+\delta \theta) = \frac{\ln(\lambda_0T_{\mathrm{int}}) + \zeta(\theta + \delta\theta)}{\alpha} = V_{\mathrm{inflc}}(\theta) + \frac{\zeta \delta \theta}{\alpha}.
        \label{eq:S_temp_shift}
        \end{equation}
        However, the robustness of the algorithms in the presence of these variations has been evaluated as presented in later sections.
        
        Device-to-device variations in $(\alpha,\kappa)$ shift $(V_{\mathrm{mid}},V_{\mathrm{inflc}},\Delta V_{\mathrm{transit}})$. Our per-pixel calibration recenters each device at $V_{\mathrm{mid}}$ and sets a local gain, after which PAOA absorbs residual mismatch into the learned variational parameters (e.g., schedules or couplings). Global reverse bias primarily scales $\lambda_0$; large changes require re-centering but were not used during PAOA runs.
        
        \section{Description of the characterization setup}
        \label{sup:setup}
        \begin{figure}[h]
            \centering
            \includegraphics[width=\linewidth]{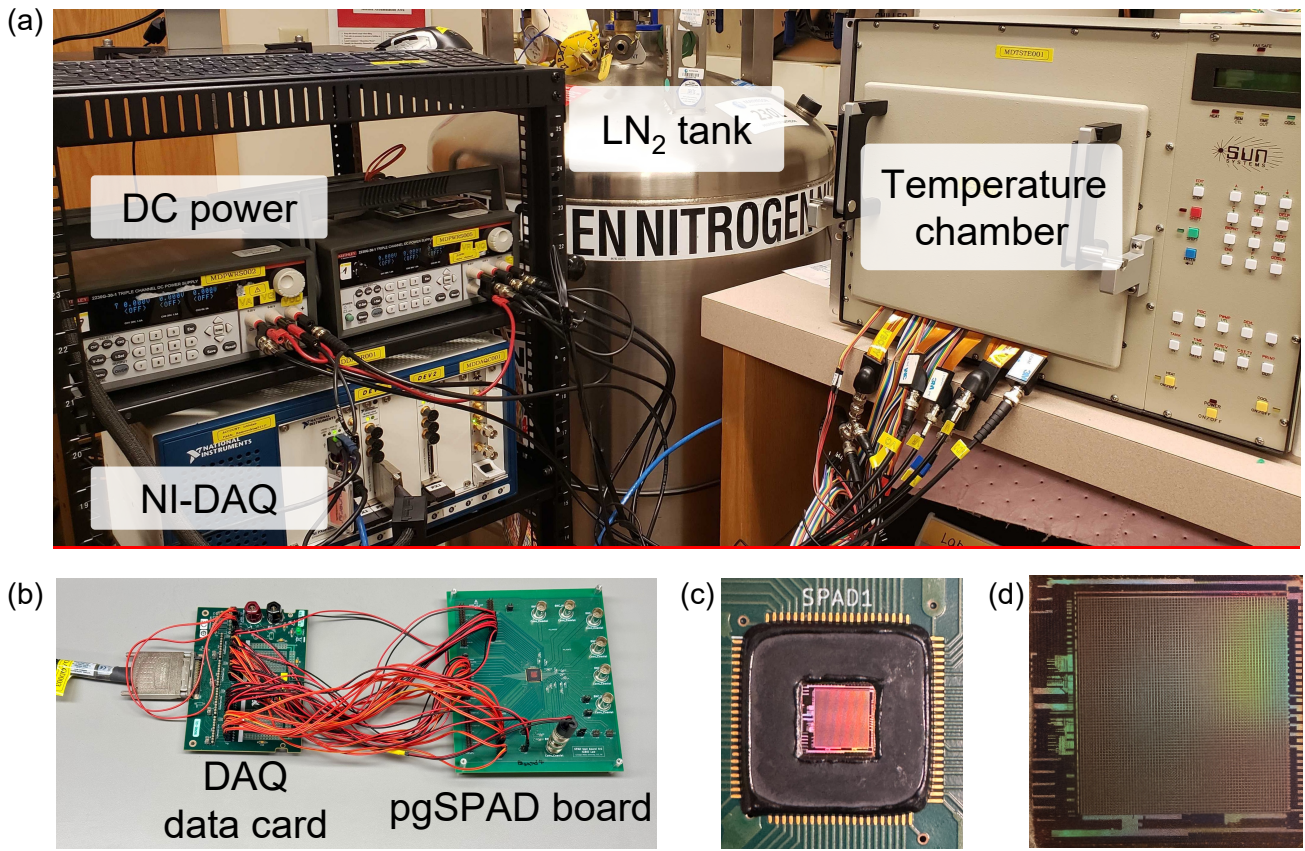}
            \caption{Experimental setup for pgSPAD characterization. (a) Temperature chamber, LN$_2$ tank, and power supplies. (b) DAQ-based timing and data collection. (c) Wire-bonded chip on PCB with epoxy-protected bonds. (d) Micrograph of the 64$\times$64 pgSPAD array with peripheral logic.}
            \label{fig:setup}
            \vspace{10pt}
        \end{figure}
        Fig.~\ref{fig:setup} (a) shows the instrumentation. A temperature chamber (EC1X, Sun Electronic Systems, Inc.) provides a temperature controlled dark environment for dark noise characterization and utilization. The temperature chamber was computer controlled to sweep and maintain target temperatures within a $\pm0.25 ^oC$. Two triple-output DC power supplies (Keithley 2231A-30-3) provided the power and the analog bias voltages to the board hosting the 64$\times$64 pgSPAD chip. A National Instruments data acquisition system (PXIe-8821) carrying a data acquisition module (NI PXIe 6547) generates timing, sets integration windows, and reads out counts as shown in Fig.~\ref{fig:setup} (b). Custom MATLAB scripts coordinate temperature sweeps and data logging. The die is wire-bonded directly to the PCB as shown in Fig.~\ref{fig:setup} (c); peripheral logic enables pixel selection and aggregated readout. A close-up view of the die is shown in Fig.~\ref{fig:setup} (d).

\section{Activation-Aware Learning with PAOA}
\label{sec:MAJ}
        
        \label{PAOA_MAJ_gate}
        \begin{figure}[htp!]
            \centering
            \includegraphics[width=0.95\linewidth]{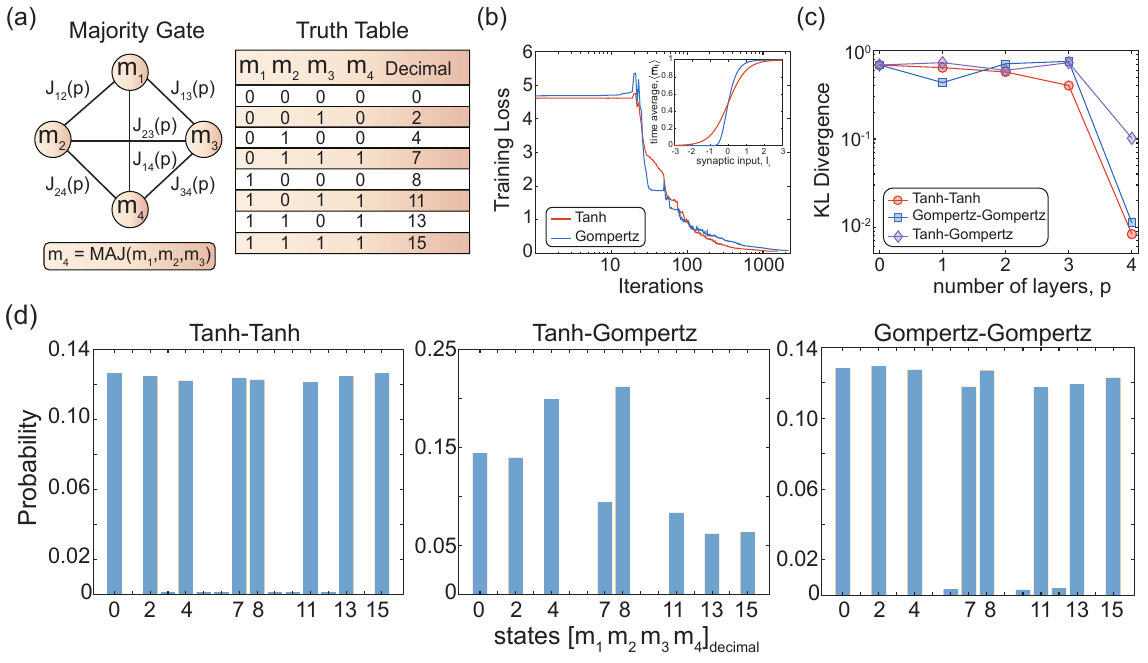}
            \vspace{-10pt}
            \caption{\footnotesize \textbf{Majority gate: demonstrating learning in PAOA under asymmetric activation.} (a)  Fully connected four-node network used to implement the majority gate $m_4$\,=\,$\text{MAJ}(m_1,m_2,m_3)$, where $m_4$\,=\,$m_1 \lor m_2$ if $m_3$\,=\,$1$ and $m_4$\,=\, $m_1 \land m_2$ if $m_3$\,=\,$0$. The table lists the eight valid input-output combinations, labeled by their decimal representation [$m_1 \ m_2\  m_3\ m_4$] using left most significant bit notation. (b) Training loss over optimization iterations for symmetric (tanh, red), and asymmetric (Gompertz, blue) activations. (c) Kullback–Leibler divergence across PAOA depth $p$ for three scenarios: training and inference with tanh (red circles), training and inference with Gompertz (blue squares), and training with tanh followed by inference with Gompertz (purple diamonds). (d) Final distributions at depth $p=4$ for the corresponding cases in (c).}\vspace{15pt}
            \label{MAJ_gate}
        \end{figure}

To illustrate the capability of PAOA to absorb activation non-idealities by learning parameters tailored to a given sampler, we solve a four-node majority gate problem using both symmetric (i.e., tanh) and asymmetric (i.e., Gompertz) activation functions. This problem serves as a tractable testbed for understanding the role of activation function asymmetry. 

The majority gate, illustrated in Fig.~\ref{MAJ_gate}a, is defined over four binary variables \( [m_1, m_2, m_3, m_4] \), where the output \( m_4 \) is given by
    \begin{equation}
        m_4 = \text{MAJ}(m_1,m_2,m_3) =  \begin{cases}
            m_1 \lor m_2, & \text{if } m_3 = 1 \\
            m_1 \land m_2, & \text{if } m_3 = 0
        \end{cases}
        \label{eq:MAJ_gate}
    \end{equation}
    
 The eight correct input-output combinations, shown in the truth table in Fig.~\ref{MAJ_gate}a, define the target set of states. The goal of the optimization is to find variational parameters that cause the final distribution to be evenly concentrated on this set. Starting from a uniform initial distribution (i.e., at $p=0$), PAOA is run for four layers (i.e., $p=4$), collects \( 10^8 \) independent MCMC samples, and optimizes the variational ansatz $\{J_{ij}(p)\}$ using the gradient-free optimizer COBYLA. The optimization terminates upon parameter convergence or upon reaching a maximum number of function evaluations. The hyperparameter values used in the simulation are listed in Table.~\ref{tab:simulation_parameters_MAJ}.
 
 Results for the training loss are shown in Fig.~\ref{MAJ_gate}b, where the two curves corresponding to the two activation functions converge to nearly identical minima. Fig.~\ref{MAJ_gate}d shows the final distributions for three scenarios: training and inference using the same activation function (i.e., ``Tanh-Tanh" and ``Gompertz-Gompertz"), and training using tanh followed by inference using Gompertz. The agreement between ``Tanh-Tanh" and ``Gompertz-Gompertz" final distributions, quantified by the Kullback–Leibler divergence in Fig.~\ref{MAJ_gate}c and qualitatively evident from the histograms in Fig.~\ref{MAJ_gate}d, demonstrates that the variational nature of PAOA can overcome activation non-idealities commonly encountered in hardware implementations due to  device variability and thermal sensitivity. We observe that inference using a different activation function than that used during training still correctly identifies the correct peaks of the target distribution, however, training the PAOA with a matched activation function results in better solution quality.

\begin{table}[htp!]
    \centering
    \caption{Majority Gate Simulation Parameters.}
        \vspace{1pc}
    \label{tab:simulation_parameters_MAJ}
    \renewcommand{\arraystretch}{1.2} 
    \setlength{\tabcolsep}{8pt} 
    \begin{tabular}{@{}ll@{}}
        \toprule
        Parameter & Value \\ 
        \midrule
        number of nodes ($N$) & 4  \\ 
        number of layers ($p$) & 4  \\ 
        update order & $\{m_1, m_2, m_3, m_4\}$\\ 
        initial parameters (${J}_{ij}(p)$) & $0.1$  \\ 
        maximum iterations ($t_{\max}$) & $5000$ \\ 
        tolerance ($\varepsilon_{\text{step}}$) & $10^{-7}$ \\ 
        number of experiments ($N_E$) & $10^8$ \\
        \bottomrule
    \end{tabular}
    \end{table}

        \section{PAOA Algorithm for SK Model}
             In this section, we explain the PAOA algorithm based on a two-schedule ansatz used to solve the SK problem. We generate 60 all-to-all SK instances sampled from a standard normal distribution, using a $50\%$-$50\%$ split between training and test datasets. Inspired by QAOA, we use PAOA with two temperature schedules, $\beta_1(p)$, and $\beta_2(p)$ are used, resulting in a total of $2p$ variational parameters. We assign schedule~1 to spins $1,\dots,N/2$ and schedule~2 to spins $N/2+1,\dots,N$, which is an arbitrary partition for SK. The training is carried out independently on each training instance, and the averaged two-schedule obtained from the training instances is then applied to the testing instances. The hyperparameter values used during training are listed in Table.~\ref{tab1:simulation_parameters}. 
             
    \label{sup:algo}
     \begin{algorithm}[htp!]
            \caption{PAOA: A Variational Monte Carlo (VMC) algorithm}
            \label{algo:PAOA}
            \SetKwInOut{Input}{Input}
            \SetKwInOut{Output}{Output}
            \SetKwFunction{FPComputer}{p-computer}
            \SetKwFunction{FPAOA}{PAOA-circuit}
            \SetKwProg{Fn}{Function}{:}{}
            \SetKwBlock{Do}{do}{end do}
        
            \Input{number of nodes $N$, number of layers $p$, number of experiments $N_{\mathrm{E}}$, maximum iterations $M$, tolerance $\epsilon_{\mathrm{step}}$, weight matrix $J$, bias vector $h$, initial two-schedule parameters $\{\beta_1(i),\beta_2(i)\}_{i=1}^p$}
            \Output{optimized double schedules $\{\beta_1(i),\beta_2{(i)}\}_{i=1}^p$}
        
            \Fn{\FPComputer{$J,h,\{\beta_1{(i)},\beta_2{(i)}\}_{i=1}^p, N, p$}}{
                initialize all spins randomly\;
                \For{$i \leftarrow 1$ \KwTo $p$}{
                    
                    \For{$j \leftarrow 1$ \KwTo $N$}{
                        \If{$j\leq N/2$}{
                            $\beta \leftarrow \beta_1{(i)}$\;
                        }\Else{
                            $\beta \leftarrow \beta_2{(i)}$\;
                            }
                        update p-bit $j$ at layer $i$ by solving Eqs.~\eqref{eq1:algo1}--\eqref{eq2:algo1}\;
                    }
                    
                }
                \Return p-bit states\;
            }
        
            \Fn{\FPAOA{$N_{\mathrm{E}}$, $J$,$h$, $\{\beta_1^{(i)},\beta_2^{(i)}\}_{i=1}^p$, $N$, $p$}}{
                \For{$n \leftarrow 1$ \KwTo $N_{\mathrm{E}}$}{
                    $\mathrm{state} \leftarrow \FPComputer({J,h,\{\beta_1^{(i)},\beta_2^{(i)}\}_{i=1}^p, N, p})$\;
                    compute the energy given by Eq.~\ref{SK_energy}\;
                }
                
                compute the average energy\; 
                \Return average energy\;
            }
        
            \While{step size $>\epsilon_{\mathrm{step}}$ \textbf{and} iterations $< M$}{
                average  energy $\leftarrow$ \FPAOA($N_{\mathrm{E}}$, $J$,$h$, $\{\beta_1^{(i)},\beta_2^{(i)}\}_{i=1}^p$, $N$, $p$)\;
                minimize average energy and get a perturbation vector ($\delta\beta_1, \delta\beta_2$) from COBYLA\;
                
                $\beta^{(t+1)}_{1} \leftarrow \beta^{(t)}_{1} + \delta\beta_1$\;
                $\beta^{(t+1)}_{2} \leftarrow \beta^{(t)}_{2} + \delta\beta_2$\;
                $t \leftarrow t + 1$\;
            }
            \Return{optimized two-schedule variational parameters}\;
        \end{algorithm}
   The classical energy function for the SK model is defined as 
        \begin{equation}
            E(\{s\}) =   -\frac{1}{\sqrt{N}} \sum_{i<j} J_{ij}s_is_j
            \label{SK_energy}
        \end{equation}
        where $\{s\}$, $N$, and $J_{ij}\sim \mathcal{N}(0,1)$ represent the system state configuration, graph size, and graph connection weights sampled from the standard normal distribution. 
        
        Using probabilistic computers, each p-bit updates its state according to  

        \begin{equation}
                I_i = \sum_j J_{ij}s_j + h_i     
                \label{eq1:algo1}
        \end{equation}
        \vspace{-1pc}
        \begin{equation}
            s_i = \text{sgn}[f\big(\beta I_i\big)-\text{rand}_{[-1,1]}] 
            \label{eq2:algo1}
        \end{equation}
        where $s_i = 2m_i-1$ denotes the bipolar p-bit state, and $J$, $h$, $\beta$, and $f(x)$ represent the problem weight matrix, bias vector, inverse temperature, and activation function (tanh for Glauber dynamics), respectively. The term $\text{rand}_{[-1,1]}$ is a uniform random variable drawn from the interval $[-1,1]$. Glauber dynamics uses the $\tanh$ activation in order to enforce ergodicity, satisfy detailed balance, and ensure convergence to the Boltzmann distribution at equilibrium, thereby minimizing the system energy. Here, however, we experiment with alternative activation function obtained by fitting the stochastic output of the pgSPAD device to the closest sigmoidal family function, which is well described by a Gompertz form. This activation is asymmetric and is given by
        \begin{equation}
            f(x) = 2\exp(-\kappa\exp(-\alpha x))-1, 
        \end{equation}
        where $\kappa$, and $\alpha$ capture device-to-device variability. As established in section.~\ref{PAOA_MAJ_gate}, optimal performance benefits from training the variational parameters 
using the same activation function employed during inference, particularly as circuit depth increases. Hence, we first sample the stochastic output of the pgSPAD device and determine the corresponding best-fit Gompertz parameters $\kappa$ and $\alpha$. Training is then performed on a CPU using this calibrated activation function with a large number of independent runs. The optimized parameters obtained from training are subsequently used for inference, with the pgSPAD serving as the probabilistic computer.

        \begin{table}[htp!]
            \centering
        \caption{Parameters used in Algorithm 1.}
            \vspace{1pc}
        \label{tab1:simulation_parameters}
        \renewcommand{\arraystretch}{1.2} 
        \setlength{\tabcolsep}{8pt} 
        \begin{tabular}{@{}ll@{}}
            \toprule
            Parameter & Value \\ 
            \midrule
            number of nodes ($N$) & 26  \\ 
            number of layers ($p$) & \{$1, \dots, 17$\}  \\ 
            initial variational parameters ($\mathbf{\beta}_1(p), \beta_2(p)$) & $1$  \\ 
            variational parameters tolerance ($\varepsilon_{\text{step}}$) & $10^{-4}$ \\ 
            maximum iterations ($t_{\max}$) & $5000$ \\ 
            number of independent experiments/runs ($N_E$) & $10^6$ \\ 
            activation function ($f(x)$) & $\tanh$ / Gompertz\\
            variability parameters $(\alpha,\kappa)$ & $(1.4,\ln2)$\\
            \bottomrule
        \end{tabular}
        \end{table}

        \end{document}